\documentclass[aps,prl,showkeys,twocolumn,amssymb,groupedaddress]{revtex4}

\usepackage{graphicx}
\usepackage{dcolumn}
\usepackage{bm}
\usepackage{ulem}

\begin{document}

\title{Relationships among detector signals recorded during events SN1987A and GW170817}

\author{N. Yu. Agafonova}
\email{Agafonova@inr.ru}
\author{A. S. Malgin}
\affiliation{Institute for Nuclear Research of the Russian Academy of Sciences, 60-letiya Oktyabrya prospekt 7a, 117312 Moscow, Russia}
\author{E. Fischbach}
\affiliation{Department of Physics and Astronomy, Purdue University, West Lafayette, IN 47907, USA}


\begin{abstract}
The temporal coincidences of events detected in four neutrino detectors and two gravitational antennas  still remains among the most puzzling phenomena associated with SN1987A. 
The coincidences form a six-hour signal approximately coincident in time with the well-known LSD signal at 2h52m UT on 23/02/1987. After 30 years of research, the characteristics and the shape of the six-hour signal have been studied quite well, but the mechanisms of its formation have not been fully understood as of yet.
Here we suggest that data obtained from another technology, radioactive decays, might provide new
insights into the origin of signals previously seen in neutrino detectors and gravity wave detectors.
On August 17, 2017, at 12h41m UT, the GW170817 signal was detected by LIGO and Virgo. At the same time, an approximately 7-hour long signal coincident with GW170817 was detected in the Si/Cl experiment on precision measurement of the $^{32}$Si half-life. We show that the Si/Cl signal is unexpectedly similar to the six-hour signal from SN1987A. In addition, we establish that the sources of the coinciding events are similar to those of the Si/Cl signal. 
To explain the surprising similarities in both signals, we present a mechanism which could in principle account for this phenomenon in terms of a local increase in the density of axionic dark matter induced by a gravity wave.
\end{abstract}


\maketitle

\section{1. Introduction}
On August 17, 2017, at 12h41m04s UTC (below we assume UTC time to be identical to UT), LIGO and Virgo gravitational detectors  both recorded a gravitational wave signal \cite{Abb17PRL}, \cite{Abb17AP} designated as GW170817. The ensuing analysis of its characteristics showed that, most likely, the signal was caused by a merger of two neutron stars (NS) with masses of approximately 1.2M$_{\odot}$ and $\sim$1.6M$_{\odot}$. The object was located in the galaxy NGC4993 of the constellation Hydra at a distance of 
40 Mps ($\sim$1.3$\times$10$^8$ light years) from the Earth, the closest for any gravitational-wave event ever recorded. A train of nearly flat gravitational waves (GW) reached the southern hemisphere of the Earth in the region of the Fiji islands. Gravitational detectors began to record the signal  approximately 30 s before the NS merger, when the GW frequency increased to the lower limits of detectors' sensitivity ranges  at about 30 Hz. The subsequent increase in frequency to $\sim$0.5 kHz (kilohertz) was accompanied 
by an increase in the GW amplitude. This ended with a peak at the very moment of NS merger that produced a time-frequency spectrogram, 
which is known as a "chirp".
This was the first time that GWs from NS merger were recorded. 
All signals detected except GW170817 (GW190425 was not marked as a significant event) were identified as coming from black hole (BH) mergers that occur over shorter times, on the order of hundreds of milliseconds.

The search for a neutrino flux from GW170817 failed; no detector capable of detecting neutrinos with energies up to $\sim$100 MeV (Borexino, Daya Bay, HALO, IceCube, KamLAND, LVD, and Super-K) has reported any neutrino signals.

\begin{figure}
\centering
  \includegraphics[width=0.50\textwidth]{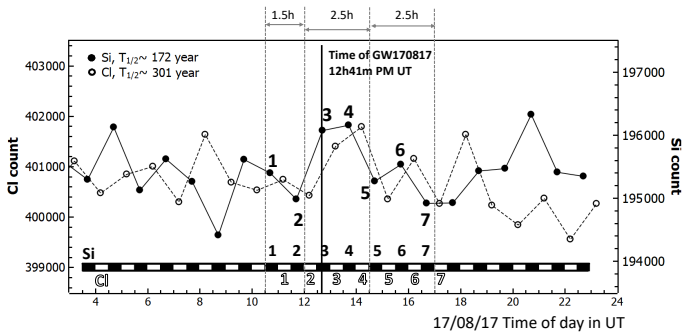}
\caption{$^{32}$Si and $^{36}$Cl counting rates in half-hour intervals (data from \cite{Fis18}). Seven pairs of correlated counts 
forming the Si/Cl signal are numbered. The horizontal black-and-white band shows the times of half-hour measurement 
intervals for Si (black bars) and Cl (white bars). The vertical dashed lines delimit parts of the Si/Cl signal based 
on Si data, and the solid vertical line denotes the GW170817 signal time}
\label{pic02}
\end{figure}

Nevertheless, in the data from \cite{Fis18}, an experiment on the  precision measurement of the $^{32}$Si half-life (T$_{1/2}\sim$172 years), a correlation between $^{32}$Si and $^{36}$Cl decay rates was observed and associated with GW170817. This correlation was observed for a $\sim$ 7 h period coincident with that of GW170817 (between point 1 and point 7 in Fig.~\ref{pic02}).

In Sec. 2, we discuss the relationship between the LIGO signal GW170817 and a correlation in the event rates in a Si/Cl decay experiment found by the authors of the experiment.
In Sec. 3 we discuss the relationship of the SN1987a neutrino events and signals in gravitational wave detectors running at the time. This was investigated in the papers cited in this section and established quantitatively in the form of the probabilities of coincidence of detector events (Sec. 3.1). One of the important points is the proof of the muon origin of the signals from the gravitational antennas in Rome and Maryland, including the signals recorded by the antennas during SN1987A (Sec. 3.2).
The main objective of our research is the analysis of experimental data taking into account the methodological features of the experiments. We show that two qualitatively different instrumental technologies appear to have seen the same astrophysical signal (Sec. 4).

To explain the surprising similarities in both signals, we present a mechanism which could in principle account for this phenomenon in terms of a local increase in the density of axionic dark matter induced by a gravity wave (Sec. 5).

\section{2. $\textbf{Si/Cl}$ signal}
\label{Sec:2} 

The experiment described in \cite{Fis18}, conducted by researchers from various US universities and research organizations, has been in progress since 2014 at Purdue University. The $\beta^-$ decay of $^{32}$Si is identified using electrons from the decay of $^{32}$P (T$_{1/2}$  $\sim$14.28 days, Q($\beta^-$) = 1709 keV) that is part of the $\beta$-decay chain $^{32}$Si $\to$ $^{32}$P $\to$ $^{32}$S, and occurs in equilibrium with $^{32}$Si. 
The decay rate of the $^{32}$Si sample is compared with the decay rate of $^{36}$Cl (T$_{1/2}$ $\sim$ 3.1$\times$10$^5$ years) that remains almost constant due to the very long T$_{1/2}$ time and, hence, is used as a calibration standard. The $^{32}$Si half-life is determined from the $^{32}$Si/$^{36}$Cl ratio that changes with time due to the decay of $^{32}$Si.

The experimental setup is based on a scintillation detector, and is protected from electromagnetic noise and supply voltage 
fluctuations. Measurements are performed under extremely stable pressure, humidity and temperature conditions \cite{Hei15}. 
A precision mechanical manipulator alternately positions samples of $^{32}$Si and $^{36}$Cl under the detector for measurements. The number of decays for each isotope is measured every hour in half-hour exposures.

The data form two time sequences, in which Cl values in accordance with the measurement method 
are shifted by 0.5 h relative to Si (Fig.~\ref{pic02}). 
In these sequences a 7-hour long segment was discovered. Within this segment the GW170817 signal is present, and Si and Cl 
decay counts change synchronously. Changes in decay rates are very small and do not exceed a fraction of a per cent.
The Pearson correlation coefficient for decays in this interval was 0.954 \cite{Fis18}.
A statistical analysis of the data covering a period 
from January 1 to October 11, 2017 showed that, firstly, fluctuations in the recorded number of decays for both Si and Cl follow the usual $\sqrt{N}$ rule for statistically uncorrelated data. Secondly, the probability of a random correlation of seven pairs of numbers (counting rates) over a seven-hour interval was no more than 4.3$\times$10$^{-4}$, 
which corresponds to a correlation coefficient of 0.954 \cite{Fis18}.

Two more seven-hour  segments (May 25 and June 15) with random correlation probabilities of no more than 4.3$\times$10$^{-4}$ 
were identified in the records spanning 9.5 months (from 01/01/17 to 11/10/17) in addition to the seven-hour signal contemporaneous with GW170817. 

On the other hand, for the full 13-month period of searching for gravitational wave signals (O1+O2 data sets), 11 signals were identified \cite{O1+O2}. Thus, the probability of a random coincidence of the Si/Cl and GW170817 signals is 2.5$\times$10$^{-5}$.

Since the probability for a rare background fluctuation to occur at a certain predetermined 
point in time (same as GW170817) is low, the authors \cite{Fis18} associated the seven-hour Si/Cl signal in August with GW170817, while those for May and June were attributed to statistical fluctuations. 
The significance of the Si/Cl signal and its connection with GW170817 is now further supported by the observation of a similar correlation signal in $^{44}$Ti and $^{60}$Co decays, the presence of which is shown in  Ref.\cite{Fis20}, and will be discussed in detail elsewhere.

As follows from Fig.~\ref{pic02}, the main features of the Si/Cl-signal for 1-hour measurement steps (set based on Si data) are as follows:\\
a)  total duration of approximately 6.5 h (7 pairs of points, from 1 to 7);\\
b) two-humped signal shape;\\
c) duration of each peak is approximately 2.5 hours.

\begin{figure}
\centering
\includegraphics[width=0.45\textwidth]{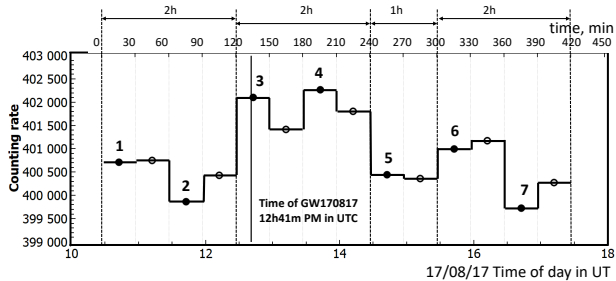}
\caption{Si/Cl signal at 0.5-hour time resolution. The number of Si counts (black circles) is normalized to the number of Cl counts (empty circles) in the range from point 1 to point 7. The upper scale is for the time elapsed since the beginning of the Si/Cl signal, and the vertical dashed lines delimit the phases of the signal. The solid black line denotes the time of the GW170817 signal}
\label{pic03}
\end{figure}

The Cl data, shifted by 0.5 h relative to the Si data as discussed above, makes it possible to increase the temporal resolution and refine the signal shape. For this, it is necessary to normalize the Si and Cl decay counts (from point 1 to point 7). 
This produces a histogram with  0.5-hour bins (Fig. ~\ref{pic03}). The first peak is then 2-hours wide (from 120 to 240 min) and is preceded by a two-hour baseline. The first peak is followed by an hour-long dip and the second peak, smaller in height and lasting {1~hour} (from 300 to 360 min), that may be accompanied by a baseline for approximately an hour (from 360 to 420 min). 
Thus, with a better time resolution, the total duration of the Si/Cl-signal increases to $\sim$ 7 h; 
both peaks become narrower, and an hour-long dip appears between them.

The duration of the Si/Cl signal (as well as the signal itself), combined with the presence of a two-hour segment preceding GW170817, is a great challenge to any interpretation based on the current understanding of the mechanisms 
for catastrophic astrophysical phenomena (mergers of BH and NS, gravitational collapses) and properties of the resulting radiations. In this regard, the authors of the Si/Cl experiment mention the LSD signal from SN1987A at 2h52m UT on February 23, 1987 \cite{Dad87} as an example of a possible precursor signal approximately 5 h ahead of the neutrino flux detection by KND \cite{Hir87}, IMB \cite{Bio87}, and BST \cite{Ale87} at 7h35m UT.

\section{3. Signals Recorded by Detectors During the SN1987A Burst}    

To begin, it is necessary to point out that the 7-second LSD signal at 2h52m36.8s UT is merely a concentration of five LSD events (pulses) located in the central part of the clearly visible six-hour long signal formed by coinciding events of multiple detectors. The signal has been extracted and studied as a result of cross-correlation analysis of the data from six detectors, carried out by various researchers between 1987 and 2016.

At the time of SN1987A, four underground detectors capable of detecting neutrinos from supernovae, and two gravitational antennas (GA) were in operation. These include: the Mont Blanc Liquid Scintillation Detector (LSD in the following) and the Baksan Scintillation Telescope (BST in the following), Cherenkov detectors KND (the Kamioka Nucleon Decay Experiment) and IMB (the Irvine Michigan Brookhaven experiment), gravitational antennas in Rome (GEOGRAV, hereinafter RGA) \cite{Bro85} and Maryland (MGA) \cite{Web69}. Characteristics of the neutrino detectors and gravitational antennas as of February 23, 1987, relevant to the present discussion, are shown in Tables \ref{tab1} and \ref{tab2}.

\begin{table*}
\caption{Characteristics of neutrino detectors}
\label{tab1}       
\scalebox{0.75}{
\begin{tabular}{lllll}
\hline\noalign{\smallskip}
 & LSD & KND & BST & IMB  \\
\noalign{\smallskip}\hline\noalign{\smallskip}
Coordinates & 45.8$^{\circ}$N~~6.9$^{\circ}$E & 36.4$^{\circ}$N~~137.3$^{\circ}$E & 42.4$^{\circ}$N~~42.7$^{\circ}$E  & 41.7$^{\circ}$N~~81.3$^{\circ}$W \\
Depth (flat overburden), m w.e. & 5200 & 2700 & 850 & 1570 \\
Altitude, m & +1380 & +370 & +1850 & $-$420 \\
Effective Volume Mass, t & 90 & 2140 & 200 & 3300 \\
Muon average energy, GeV & 385 & 260 & 140 & 190 \\
Muon counting rate, h$^{-1}$ & 4.2 & 1.3 $\times$ 10$^3$ & $\sim$ 10$^5$ & 9.7 $\times$ 10$^3$ \\
Energy threshold, MeV & 5 (int.) 7 (ext.) & 7.5 (20 hits) & 10 & 25 (24 hits) \\
Energy resolution & 25\%; 10 MeV,e,$\gamma$ & 22\%; 10 MeV,e$^-$ & 20\%; 12--20 MeV& 25\% \\
Background counting rate, h$^{-1}$ & 43 & 86 & 1.2 $\times$ 10$^2$ &  8.1 $\times$ 10$^3$ \\
Accuracy of detector clock & $\pm$ 2 ms & $\pm$ 60 s & $-$54 s, +2 s & $\pm$ 50 ms \\
Resolution time (no worse) & 0.2 $\mu$s & 0.2 $\mu$s & 0.2 $\mu$s & 0.2 $\mu$s \\
\noalign{\smallskip}\hline
\end{tabular}
}
\end{table*}

\begin{table*}
\caption{Characteristics of gravitational antennas in Rome (RGA) and Maryland (MGA)}
\label{tab2}
\scalebox{0.8}{
\begin{tabular}{lll}
\hline\noalign{\smallskip}
 & RGA & MGA \\
\noalign{\smallskip}\hline\noalign{\smallskip}
 Coordinates &  42.0$^{\circ}$N~~12.5$^{\circ}$E & 39.0$^{\circ}$N~~77.0$^{\circ}$W\\
Altitude, m & +40 & +20 \\
Substance, form & Al, cylinder, L=3.0m, D=0.6m             & Al, cylinder, L=1.55m, D=1.0m\\
Mass, t & 2.3 & 3.1 \\
Eigen vibration frequency, Hz & 858 & 1660 \\
Orientation & 29$^{\circ}$ E-W & 0$^{\circ}$ E-W \\
Accuracy of detector clock & $\pm$ 0.1 s & $\pm$ 0.1 s \\
Effective temperature & 29 K & 31 K \\
\noalign{\smallskip}\hline
\end{tabular}
}
\end{table*}

For the first time the possibility of correlations between the LSD and RGA events recorded during SN1987A, has been pointed out in \cite{Ama87}. The relevant LSD event was a pulse corresponding to an energy release of E$_{\mathrm{th}}$ $\geq$5 MeV in the detector. The RGA (and MGA) event was the antenna excitation energy (temperature) that was measured every second and expressed in Kelvins (K). When analyzing the RGA antenna events in the vicinity of $\pm$30 s of the LSD signal at 2h52m37s UT, events whose energy significantly exceeded the average antenna excitation energy were found. 
This phenomenon has initiated research efforts focused on cross-correlations of events of all six detectors over the longest time intervals possible. The methodology for studying correlations (event coincidences) has been described in detail in \cite{Agl89}, \cite{Ama89}, \cite{Chu89}, \cite{Agl91A}, \cite{Agl91B}.

\begin{figure}
\centering
  \includegraphics[width=0.35\textwidth]{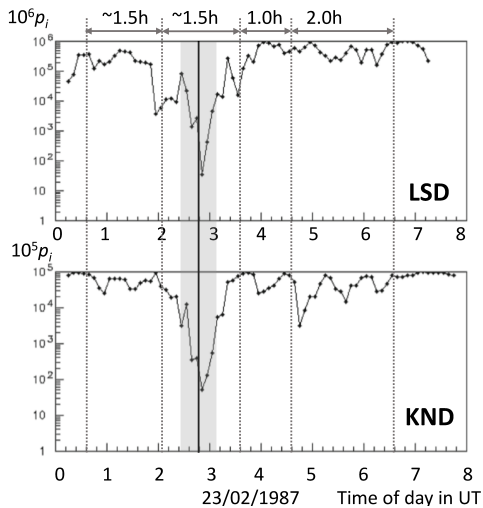}
\caption{Correlation curves obtained at  $\Delta t = \pm$ 0.5 s, $\Delta T$ = 0.5 h, $t_{\mathrm{sh}}$ = 0.1 h  for the LSD--RGA/MGA and KND--RGA/MGA setup combinations. The ordinate scale represents the probability $p_i$ scaled up 10$^6$  for the LSD--RGA/MGA and 10$^5$ for KND--RGA/MGA curve.}
\label{pic_Corr_LSD-KND}
\end{figure}

The most detailed correlation analysis, covering the range from 0h00m to 8h00m UT on 23/02/1987, has been reported in \cite{Gal16}. One of the important results of this work are the correlation curves for the LSD-RGA/MGA and KND-RGA/MGA setup combinations (Fig.~\ref{pic_Corr_LSD-KND}). 
For these combinations, $p_i(t)$ is the probability of a triple coincidence of their background pulses in the number $N_{\mathrm{cor}}$ observed within the time window $\Delta T$. A method for calculating $p_i$ of triple coincidences of setup events within the interval $\Delta t = \pm$0.5 s has been described in \cite{Ama89} and \cite{Agl91A}, from which the relation $N_{\mathrm{cor}} \propto 1/p_i$ follows. A good illustration of how $N_{\mathrm{cor}}$ and $p_i$ are related may be found in Fig. ~\ref{pic02} of \cite{Chu89} that provides an analysis of correlation for the 
LSD and BST setups (double coincidences). In the analysis, the same parameters were used as in the determination of triple coincidences: coincidence time $\Delta t = \pm$0.5 s, time window $\Delta T$ = 1 h, window shift along the time axis $t_{\mathrm{sh}}$ = 0.1 h.

The correlation curves for different combinations of setups \cite{Agl89}, \cite{Ama89}, \cite{Chu89}, \cite{Agl91A} have the same feature: a large two-hour long peak from 1h30m to 3h30m.
In combinations LSD-RGA/MGA and KND-RGA/MGA, it is followed by a one-hour dip to $\sim$ 4h30m, and a group of small peaks up to $\sim$ 6h30m. Due to the similarity of the correlation curves for different combinations of detectors, we have combined them under a common label 6D/6h signal that reflects the number of detectors (6D) and duration of the correlation region (6h) in the vicinity of the LSD signal at 2h52m UT.
 
The analysis of correlations leads to the following conclusions:\\
(a) the existence of time correlations of the events of all detectors in the six-hour neighborhood of 
    the LSD signal is an experimental fact supported by statistics;\\
(b) correlations are observed on an intercontinental scale, and are independent of experiment conditions 
    and signal detection techniques;\\
(c) correlations are observed from 0h30m to 6h30m UT on 23/02/1987, and the number of coincidences 
    is the greatest between 1h45m and 3h45m;\\
(d) time shifts applied to KND (+7.7 s) and BST ($-$29 s) events to make adjustments to KND and BST 
    clocks lead to the highest number of coincidences between their events and the events of other detectors;\\
(e) no correlations within a one-hour vicinity of the KND--IMB--BST neutrino signal at 7h35m UT 
    on 23/02/1987 in the data from all detectors were found.

\subsection{3.1 Origins of 6D/6h Signal Events in Particle Detectors LSD, KND, BST, IMB}  

The origin of the coincident events of the detectors will be estimated based on a comparison of the average values of their amplitudes with the average background pulse amplitudes within the range from 1h45m to 3h45m (Table \ref{tab3}), taking into account the distribution of events throughout a detector.

Background pulses of LSD and KND detectors, which had low detection thresholds (Table \ref{tab1}), should be grouped by energy near the threshold due to the steeply decreasing power spectrum of the background $N_{\mathrm{bg}} \propto E^{-\alpha}$, 
where $\alpha \approx$ 3 is within the energy range of 5 to 10 MeV \cite{Agl86}. The background pulses come mainly from the radioactivity of detector materials and surrounding rock.

The main background of the BST setup (single 10 to $\sim$50 MeV pulses picked up by counters) at a threshold of 10 MeV is produced by muons ($\sim$10$^5$ $\mu$/h), causing one-time signals in single counters, and noise pulses from photomultipliers \cite{Ale89}.

The values given in Table \ref{tab3} show that the average amplitudes of all LSD events correlating with KND and BST events, and KND events correlating with LSD events, are near the detection threshold, taking into account the energy resolution of 20--25\% (Table \ref{tab1}). 
In addition, the mean energy $\bar{E}$ = 6.7 MeV of the LSD signal at 2h52m (5 pulses) is almost the same as the mean energy $\bar{E}_{\mathrm{bg}}$ = 6.8 MeV of LSD background events.

On the basis of the BST--LSD correlation analysis in \cite{Chu89}, we can draw a conclusion concerning the origin of the correlating events of the BST detector. The results of the analysis show that the spatial distribution of the BST events correlating with LSD events (in the period of 1h45m to 3h45m) is consistent with the distribution of background pulses within the BST setup. Herewith, the number of correlating events in 650 outer counters (7 events total) is 3.2 times greater than the number of events in 1,200 inner counters (4 events), when expressed on per-counter basis. This value practically coincides with the ratio of 3.1 per counter per hour for background pulses. The energy of the correlating pulses $E_{\mathrm{cor}}$ = 23.6 MeV (Table \ref{tab3}), which significantly exceeds both the 
threshold energy ($E_{\mathrm{th}}$ = 10 MeV) and the background pulse energy $E_{\mathrm{bg}}$, suggests a significant 
contribution made by muon crossings of only one counter to the total number of correlating events.

The origin of IMB events is obvious: given the 20 MeV energy threshold of the detector, all of them are produced by muons and products of their interactions.

The energy distributions of the background pulses and correlating events in the LSD, KND, and BST experiments plotted in Fig. 11, 13, and 14 in \cite{Agl91A} show that, in general, these distributions in all setups are consistent with one another. These facts suggest that all correlating events in the LSD, KND, BST, and IMB detectors were the physical background pulses, not only within the time interval from 1h45m to 3h45m, but also for the entire duration of the 6D/6h signal.

\begin{table}
\caption{Background $N_{\mathrm{bg}}$ and correlating $N_{\mathrm{cor}}$ events of the LSD, KND, BST detectors  in the period of 1h45m to 3h45m UT 23/02/87}
\label{tab3}
\scalebox{0.6}{
\begin{tabular}{l|c|c|c}
\hline\noalign{\smallskip}
Characteristics of all  & \multicolumn{3}{c}{Characteristics of detector events correlating with events of other detectors} \\
  detector events                             & LSD & KND & BST \\
\noalign{\smallskip}\hline\noalign{\smallskip}
LSD                                     &             ---                   & $N_{\mathrm{cor}}$=9 \cite{Saa10},  & $N_{\mathrm{cor}}$=14 \cite{Saa10},  \\ 
$N_{\mathrm{bg}}$=96,  &                                 & $\overline E_{\mathrm{cor}}$=6.8 MeV & $\overline E_{\mathrm{cor}}$=7.2 MeV  \\
\noalign{\smallskip}\hline
KND                                     & $N_{\mathrm{cor}}$=9 \cite{Saa10},  & ---                           & $N_{\mathrm{cor}}$=10 \cite{Agl91A} \\
$N_{\mathrm{bg}}$=191, $\overline E_{\mathrm{bg}}$=7 MeV  &  $\overline E_{\mathrm{cor}}$=9.2MeV    &                             &                           \\
\noalign{\smallskip}\hline
BST                                     & $N_{\mathrm{cor}}$=14 \cite{Saa10},  & $N_{\mathrm{cor}}$=10 \cite{Agl91A}   & --- \\
$N_{\mathrm{bg}}$=232, $\overline E_{\mathrm{bg}}$=15 MeV & $\overline E_{\mathrm{cor}}$=23.6 MeV    &                             &   \\
\noalign{\smallskip}\hline
\end{tabular}
}
\end{table}

\subsection{3.2 Origin of Events in Gravitational Antennas RGA and MGA}  

The solid-state resonant antennas RGA and MGA (Table \ref{tab2}) were operated indoors, at room temperature,  near sea level \cite{Bro85}, \cite{Web69}. 
The antenna excitation energy was determined within a set time interval; for RGA the interval was 1~s, and for MGA is was 0.1 s, followed by summation to 1 s. Thus, any event in the antenna reflects its energy state within a 1-second interval. 
Each event has an almost constant noise (thermodynamic) component, which is in equilibrium with room temperature and noise of the electronics \cite{Bon78}.

The sequences of 55 events received in the RGA and MGA experiments between 2h51m56s and 2h52m51s UT on 23/02/1987 are  characterized by average temperatures of $\langle \textbf{T}_R \rangle$ = 29 K for RGA and $\langle \textbf{T}_M \rangle \approx$ 30 K for MGA (\cite{Agl89}).

\begin{figure}
\centering
   \includegraphics[width=0.50\textwidth]{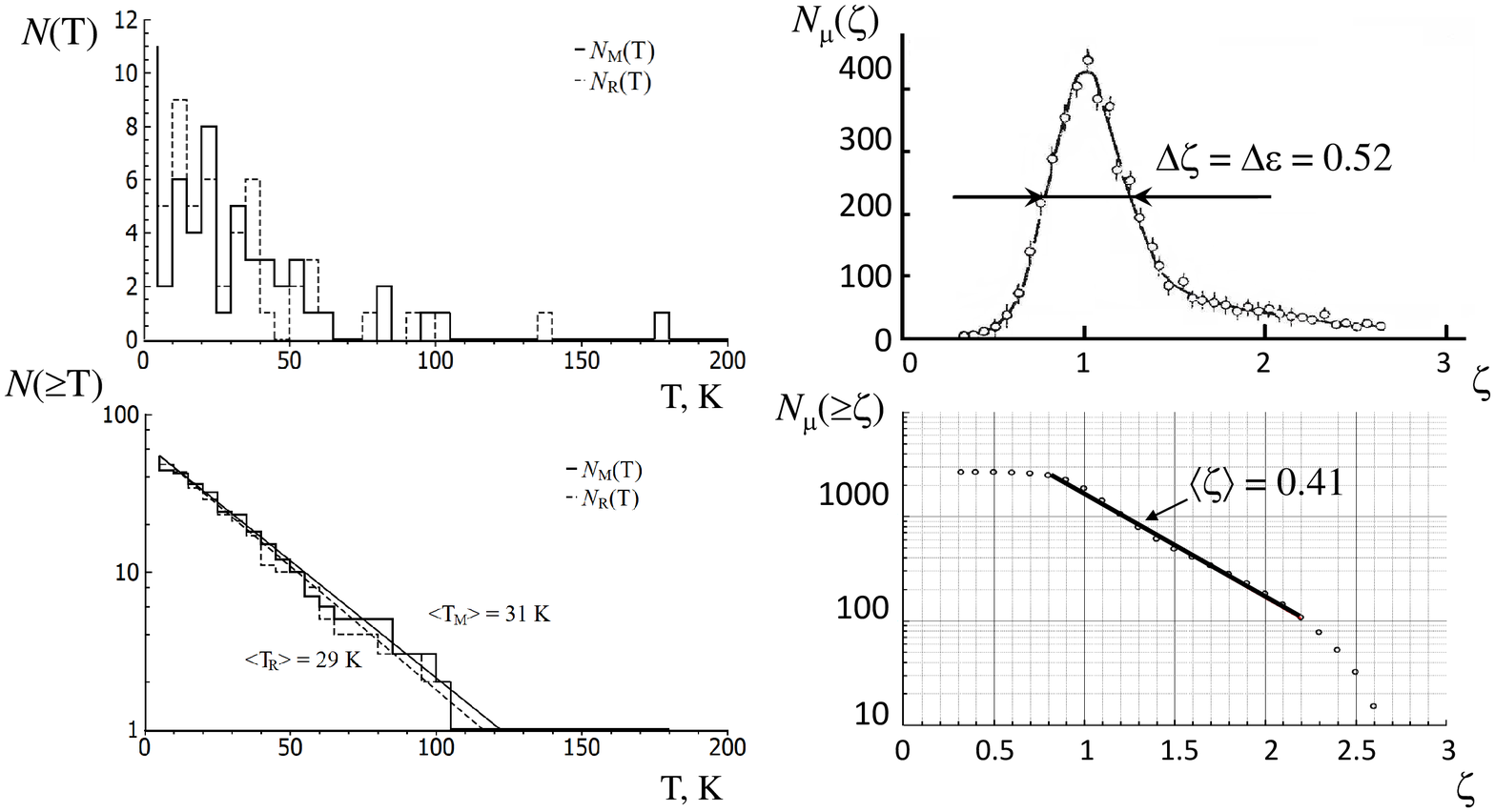}
\caption{Left: Differential and integral energy distributions of 55 RGA and MGA events in the vicinity of the LSD signal at 2h52m UT 23/02/1987. Right: Differential spectrum of muon ionization losses $N_{\mu}(\zeta)$, measured at sea level and its integral $N_{\mu}(\geq \zeta)$. 
The abscissa scale $\zeta$ represents the ratio $\zeta$=$\varepsilon$/$\varepsilon_{\mathrm{pr}}$, 
where $\varepsilon_{\mathrm{pr}}$ is the probable loss of energy; and the ordinate values are arbitrary units.
}
\label{pic_MGA_RGA}
\end{figure}

Let us consider the energy characteristics of RGA events using the differential and integral amplitude distributions shown in Fig.~\ref{pic_MGA_RGA}. 
Despite the conclusion made in \cite{Ezr70} on  the lack of influence of extensive air shower particles on MGA (and, therefore, on RGA), one cannot exclude the possibility of the formation of the distribution shown in Fig.~\ref{pic_MGA_RGA} under the action of atmospheric muon flux. 
This follows since the authors of \cite{Ezr70} have studied coincidences between gravitational antenna signals and  instantaneous antenna intersections by showers, while every RGA (and MGA) event was produced by the total excitation energy of the antenna over one-second period.

When positioned horizontally, the RGA was impacted by 290 muons per second on average:
$\bar{N}_{\mu}$ = $S I_{\mu} F$ = 1.8 m$^2 \times$130 ${\mu}{\cdot}$m$^{-2}\cdot$s$^{-1} \times$1.24 = 290 $\mu\cdot$s$^{-1}$;
where $S$ is a horizontal cross-section of the RGA, $I_{\mu}$ is a total muon flux at sea level, $F$ 
is a geometric factor taking into account the RGA shape and the angular muon distribution 
 $I_{\mu}(\theta) \propto cos^2 \theta$.

Within the framework of the hypothesis of the muon origin of the events, the histogram $N(\geq\textbf{T})$ for a detector temperature $\textbf{T}$ shown in Fig.~\ref{pic_MGA_RGA} is shaped by fluctuations in the total energy release  by muons ($\varepsilon _{\mu}^{\mathrm{tot}}$) crossing GA over a one-second interval:
 $\varepsilon _{\mu}^{\mathrm{tot}}$ = $N_{\mu}\varepsilon \bar{l}_{\mu}$, where $N_{\mu}$ is the number of muons crossing RGA per second; $\varepsilon$ is specific energy loss of muons MeV$\cdot$(g/cm$^2$)$^{-1}$; and $\bar{l}_{\mu}$=$2RL/(R+L)\times\rho_{\mathrm{Al}}$ =0.55m ($R$=0.3m, $L$=3m) $\times$ 2.7 g/cm$^3$  is the mean path of muons crossing GA.
Given the mean ionization loss $\bar{\varepsilon} \sim$ 2 MeV(g/cm$^2$)$^{-1}$, we find the total energy release of muons, $\varepsilon_{\mu}^{\mathrm{tot}} \approx$ 86 GeV.

The integral distribution $N_{\mu}(\geq \zeta)$ and the integral distribution RGA-events $N_R(\geq\textbf{T})$, MGA-events $N_M(\geq\textbf{T})$ have the  same features:\\
- Energy (temperature) values for almost all events (53 out of 55) are between 0 and 100 K, and their distribution agrees with the exponent $\mathrm{exp}(-\textbf{T}/\langle \textbf{T} \rangle)$; In the range of $\Delta\textbf{T}$ = 100K (from 0 to 100K), the $\textbf{T}/\langle \textbf{T} \rangle$ ratio decreases by a factor of 3.3 at $\langle \textbf{T}_R \rangle$ = (29$\pm$1)K and $\langle \textbf{T}_M \rangle$ = (31$\pm$4)K $\approx$ 30K.\\
- There is an exponential segment of ionization loss distribution with the exponent $\langle \zeta \rangle$ = 0.41 in the range $\Delta\zeta$ = 1.35 (from 0.85 to 2.2); the vast majority (about 96\%) of all events $N_{\mu}$ for the entire interval $\zeta\geq$ 0.85 is within this range; the $\Delta\zeta/\langle\zeta\rangle$ ratio in the range $\Delta\zeta$ = 1.35 decreases 3.3 times. 

Therefore, the temperature range $\Delta\textbf{T}$ = 0 to 100K corresponds to  $\Delta\zeta$ = 1.35; so, we can determine the coefficient $k = (\Delta\textbf{T}/\Delta\zeta)$ = 74 that connects  $\langle\zeta\rangle$ and  $\langle\textbf{T}_R\rangle$: $\langle\textbf{T}_R\rangle$= $\langle\zeta\rangle\cdot k$ = 0.41$\times$74 = 30 K.

The equality $\langle\textbf{T}_R\rangle = \langle\textbf{T}_M\rangle$ is a consequence of the muon origin of the antenna signals. 
This is due to the fact that, firstly, the total energy release of a large number of muons crossing the antenna during 1 second is proportional to its mass, regardless of its size and spatial orientation, and secondly, the heat capacity of an antenna is also proportional to its mass. Therefore, if characteristics of the muon flux are  identical (RGA and MGA were located at almost the same altitude, see Table \ref{tab3}), the antennas made of the same material should display the same temperature characteristics associated with muon energy release.

\section{4. Relationship between the $\textbf{Si/Cl}$ and $\textbf{6D/6h}$ signals}             

\begin{figure}
\centering
  \includegraphics[width=0.30\textwidth]{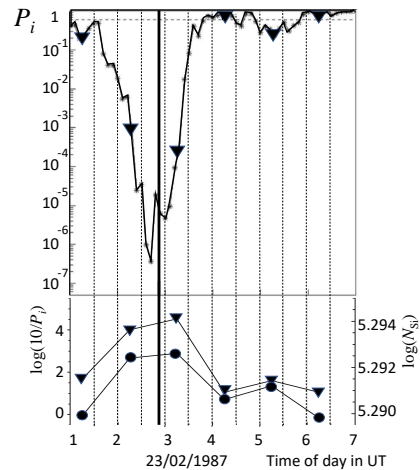}
\caption{Correspondence between signals 6D/6h and Si/Cl.
         Top panel: probability $P_i$ of similarity of correlation curves for detector combinations LSD--RGA/MGA and KND--RGA/MGA 
         at $\Delta t$ = $\pm$ 0.5 s, $\Delta T$ = 1 h, $t_{\mathrm{sh}}$ = 0.1 h. The horizontal dashed line indicates the expected $P_i$ 
         value in the absence of correlation.
         Bottom panel: shape of signals 6D/6h (triangles, left scale) and Si/Cl (circles, right scale, Si data) when
         combining the beginning of the signals. A vertical solid
         line denotes the LSD signal time 2h52m UT, a vertical dashed line indicates the relative GW170817 time}
\label{pic_Pi}
\end{figure}

The correlation curve shown on the top panel in Fig.~\ref{pic_Pi} 
(plot derived from Fig.~\ref{pic02} of \cite{Gal16}), is a graphic presentation of the 6D/6h signal at the time window 
$\Delta T$ = 1 h. The curve represents the probability $P_i$ that the correlation 
curves $p_i(t)$ for two different setup combinations (LSD--RGA/MGA and KND--RGA/MGA, Fig.~\ref{pic_Corr_LSD-KND}) are similar to each other, 
that is, they have the same probability for a fixed time $t$ (RGA/MGA are coincidences of the events of 
RGA and MGA antennas with amplitudes above a certain value \cite{Ama89}, \cite{Agl91A}).

The main characteristics of the 6D/6h signal mentioned above (total duration of about 6 h, the presence of two peaks, and the ratio of their durations) reveal an unexpected similarity with the Si/Cl signal. 
This  similarity between Si/Cl and 6D/6h signals is illustrated by the bottom panel in Fig.~\ref{pic_Pi}. The ordinate scale on the left is defined by the function $N_{\mathrm{cor}} = log(10/P_i)$, where $P_i$ are the probability values at the selected points of the curve on the top panel in Fig.~\ref{pic_Pi} that match the interval (1 h) between measurements and the their duration (0.5 h) in the Si/Cl experiment. The right scale represents the value of $log N_{\mathrm{Si}}$, where $N_{\mathrm{Si}}$ 
is the number of Si decays at points 2 to 7 in Fig.~\ref{pic02}; the number $N_{\mathrm{Si}}$ at point 2 is assigned to the time 1h30m of the 6D/6h signal. As indicated above, the interval between measurements for both curves was 1~h. Analysis with a better time resolution at 0.5 h intervals (Fig.3 in \cite{Gal16}) allowed us to refine the shape of the 6D/6h signal as follows: beginning at $\sim$ 0h40m, duration of the first peak of about 1.5 h (from 2h00m to $\sim$ 3h30m); then one-hour long dip to $\sim$ 4h30m, followed by a group of peaks during $\sim$2 h. The total signal duration is about 6 hours. 

The accuracy of time bounds determination is $\pm$ 15 min. 
The time span of the central coincidence region in this picture agrees with the range of 1h45m to 3h45m in 
\cite{Agl89}, \cite{Agl91A} within the margin of error.

\section{5. Connection of $\textbf{Si/Cl}$ and $\textbf{6D/6h}$ Signals with Gravitational Waves}  
 
We can establish a connection between the Si/Cl and 6D/6h signals if we assume that the 6D/6h signals arise from the response of the local axionic dark matter distribution to a gravity wave. 

By hypothesis, axionic dark matter (DM) can directly interact with nuclei, and hence the modified DM distribution can in principle modify nuclear decay rates locally. This mechanism thus provides a natural explanation for time-varying nuclear decay rates as a connection between the Si/Cl and 6D/6h signals.

Non-statistical variations in the intensity of radioactive decays have been reported repeatedly, most recently in \cite{Fal01}, \cite{Jav10}, \cite{Jen09}, \cite{Jen10}, \cite{Par11}, \cite{Moh16}. One of the first reliable indications of the possibility of variations (seasonal) has been obtained in an experiment aimed at measuring the half-life of $^{32}$Si in 1982 -- 1985 \cite{Alb86}. The mechanism of these variations remains unknown.

At the time of the Si/Cl and 6D/6h signals, neutrino detectors did not detect any neutrino fluxes. Other than neutrinos from known radiations, gravitational waves would be the only other possible cause of these signals. Therefore, we consider the possibility of coupling Si/Cl and 6D/6h signals with gravitational radiation. Given the temporal coincidence of the Si/Cl signal with GW170817, along with the Si/Cl and 6D/6h coupling with GW, we consider the effect of gravitational waves on the probability of radioactive decays.

We now demonstrate that the above similarity of the Si/Cl and 6D/6h signals can be explained in terms of a mechanism that is supported by the axion-based model of dark matter Ref.\cite{7a}.

Following Ref.\cite{7a} we consider a generic axion model where ``axion'' denotes a light pseudoscalar (or scalar) particle with a mass in the range of $\sim$10$^{-6}$eV. On cosmological scales, such particles behave as non-relativistic candidates for dark matter. As noted in Ref.\cite{7a}, an oscillating pressure arising from axions can induce oscillations in the gravitational potential, which could be detected in principle. In what follows we demonstrate that the preceding mechanism can be inverted to establish that a gravity wave arising from SN1987A or GW170817 can lead to a fluctuation in the local axionic field. This fluctuation can, in turn, couple to the nuclei of radioisotopes such as $^{32}$Si and $^{36}$Cl, and perturb them in such way as to produce the fluctuations observed in the respective decay rates.

Returning to Ref.\cite{7a} we consider a class of theories in which the action $S$ is given by
\begin{equation} 
S=\frac{1}{2}\int d^4x\sqrt{-g}[R+f(R)]+S_m
\end{equation}
where $R$ is the Ricci scalar, $f(R)$ is a model-dependent function of $R$, and $S_m$ is the matter field arising from 
axions. To illustrate how this formalism applies to the present work, we follow Ref.\cite{7a} and consider a model in which 
$f(R) = R^2/6M^2$, where $M$ is a constant mass scale. The field equation arising from Eq.(1) can be solved to express 
the Ricci scalar $R$ in terms of the dark matter density $\rho_{DM}$, as was done in Ref.\cite{7a}. However, for present 
purposes we are interested in how a changing gravitational field arising from SN1987A can influence the local dark 
matter density $\rho_{DM}$, through a time-dependent contribution from the Ricci scalar. From Ref.\cite{7a} this 
time-dependence is given by

\begin{equation} 
\rho_{DM} =R(t)\frac{1-(2m/M)^2}{[1-(2m/M)^2]+3\cos(2mt)},
\end{equation}
where $m$ is the axion mass. We note from Eq.(2) that a time-dependence of $\rho_{DM}$ can arise from both a change in the gravitational field given by $R(t)$, and an explicit time-dependence of the axionic field arising from $cos(2mt)$.

The preceding discussion is a model of how a time-dependent gravitational field arising from SN1987A or GW170817 
could produce the time-dependent signals discussed above, through changes in the local axionic dark matter density $\rho_{DM}$. Axionic dark matter can influence unstable nuclei through a variety of mechanisms, particularly those with a relatively small Q-value. As shown in Ref.\cite{Fis09}, a perturbation as small as 50 eV would induce a change of $\pm$1\% in tritium decay, which is typical of changes detected in other radionuclides, undergoing $\beta$-decay. For $\alpha$-decays, the results can be even more dramatic. As will be shown elsewhere \cite{37}, an increase in the Q-value of $^{222}$Rn by a few eV can change its decay rate by several orders of magnitude. These observations raise the prospect of using fluctuations in $\alpha$- and $\beta$- decays to study astroparticle phenomena, as well as a means of predicting solar storms.

\section{6. Conclusion}
\label{sec:6}

We have discussed effects with manifest themselves against the background of weak statistical fluctuations in the rates of decay of nuclei in samples. 
In two samples the effect appears as a decay synchronization, rather than a change in the overall decay rate. An indication of the conservation of the decay constant in the presence of 
short-term (seasonal) variations in the decay rate follows from the precision measurement of the lifetime of the long-lived $^{32}$Si isotope \cite{Alb86}.

The connection of the Si/Cl and GW170817 signals is confirmed by the extremely low probability of their random coincidence. This relationship indicates the possibility of the influence of GW on the probability of decays by means of a fluctuation in the local axionic field.

On the other hand, the existence of a 6D/6h signal is an established experimental fact.
Assuming that the 6D/6h signal is associated with the SN1987A burst, and based on the similarity of the shape and origin of the Si/Cl and 6D/6h signals, we can conclude that the 6D/6h signal is caused by the GW from the merger of the compact binary mass system.                 

Since no neutrino radiation was detected in the vicinity of the LSD signal at 2h52m, we propose that GWs were emitted by a system of two fragments of the star iron core that had disintegrated at the initial stage of collapse. Fragments of the precollapsar were losing energy mainly due to GR and then, after the merger, collapsed into NS, producing the neutrino burst recorded at 7h35m UT.                                                             
A similar scenario following the disintegration of the newborn neutron star and subsequent merging of its fragments is discussed in \cite{Ree74}, \cite{Ims95}. 

The possibility of inducing a signal which reaches the Earth depends on the distance to the source, the inclination of the binary mass system rotation plane relative to the direction to the Earth, and the GR energy. The availability of 
the Si/Cl signal shows that, in the case of GW170817, a combination of these conditions and the sensitivity of the Si/Cl experiment contributed to detection, despite the distance to the galaxy 
NGC4993, which is three orders of magnitude greater than the distance to SN1987A. It is difficult to say anything about the sensitivity of other detectors that could have detected weak indications of variations in the radioactive background at the time of the GW170817 signal. 
This is because the coincidences of the background events in the detectors were not analyzed in the temporal vicinity of the signal according to the method employed in \cite{Ama87}, \cite{Agl89}, \cite{Ama89}, \cite{Chu89}, \cite{Agl91A}, \cite{Agl91B}, \cite{Gal16}.                                 

The sequence of events considered above, which admits the action of gravitational field variations on the probability of radioactive decays, indicates the possibility of an additional channel for obtaining information about catastrophic astrophysical phenomena. 

We have now demonstrated that signals from catastrophic phenomena such as SN1987A and GW170817 can also be detected via synchronous changes in locally measured radioactive decay rates, along with traditional technologies such as neutrino detectors and gravity wave detectors. Given the availability of a large number of ${\alpha}$-decay and ${\beta}$-decay radioisotopes, with a wide variety of half-lives and Q-values, it seems likely that suitable candidates can be found to constitute a new useful detection system.


We conclude this discussion by noting that one implication of the present work is that the similarity of the SN1987A and GW170817 signals lends further support to currently popular models of dark matter based on light axions.

\paragraph{Acknowledgements}
The work was carried out with partial support from the grant of the Russian Science Foundation 18-02-00064 and the program of basic research of the Presidium of the Russian Academy of Sciences "Physics of fundamental interactions and nuclear technologies". 
The authors wish to thank Dennis Krause, Carol Scarlett, and Belvin Freeman for helpful discussions.

~


\begin{thebibliography}{99}
\bibitem{Abb17PRL}
B.~P.~Abbott, R.~Abbott, T.~D.~Abbott, F.~Acernese {\em et al.}, 
Phys. Rev. Lett. {\bf 119} 161101 (2017)

\bibitem{Abb17AP}
B.~P.~Abbott  {\em et al.},
ApJL {\bf 848} L12(2017) 

\bibitem{Fis18} 
E.~Fischbach {\em et al.},
Astroparticle Physics {\bf 103} 1 (2018), arXiv:1801.03585

\bibitem{Catalog} 
LIGO Scientific Collaboration and Virgo Collaboration, 
https://www.ligo.org/detections/O1O2catalog.php

\bibitem{Hei15} 
J.~Heim, 
{\it The determination of the half-life of 32Si and time varying nuclear decay} Ph.D dissertation, Purdue University (2015) 

\bibitem{O1+O2} 
Gravitational Wave Open Science Center 
https://www.gw-openscience.org 

\bibitem{Fis20} 
E.~Fischbach, D.~E.~Krause, and M.~Pattermann,
{\it Comment on "Testing claims of the GW170817 binary neutron star inspiral affecting $\beta$-decay rates"} arXiv:2003.00092

\bibitem{Dad87} 
V.~L.~Dadykin {\em et al.}, 
JTEP Letters {\bf 45}, Issue 10 593 (1987) (Pis'ma Zh. Eksp. Teor. Fiz. 45, N10 (1987) 464-466.); 
M.~Aglietta {\em et al.} 
Europhys. Lett. {\bf 3} 1315 (1987)

\bibitem{Hir87} 
K. Hirata {\em et al.},
Phys. Rev. Lett. {\bf 58} 1490 (1987) 

\bibitem{Bio87}
R.~M.~Bionta {\em et al.}, 
\emph{Phys. Rev. Lett.} {\bf 58} 1494 (1987)  

\bibitem{Ale87} 
E.~N.~Alekseev {\em et al.}, 
JETP Lett. {\bf 45} 589 (1987)

\bibitem{Bro85} 
T.~Bronzini, S.~Frasca, G.~Pizzella {\em et al.}, 
Il Nuovo Cimento C {\bf 8} 300 (1985

\bibitem{Web69} 
J.~ Weber,  
Phys. Rev. Lett. {\bf 22} N24 1320 (1969)

\bibitem{Ama87} 
E.~Amaldi, P.~Bonifazi, M.~G.~Castellano, E.~Coccia {\em et al.}, 
EPL (Europhysics Letters) {\bf 3} N12 1325 (1987)

\bibitem{Agl89} 
M.~Aglietta {\em et al.}, 
Il Nuovo Cimento C {\bf 12} (1989) 75-103

\bibitem{Ama89} 
E.~Amaldi {\em et al.}, 
Texas Symposium on Relativistic Astrophysics, (14th, Dallas, TX, Dec. 11-16, 1988) {\bf 571} 561 (1989)

\bibitem{Chu89} 
A.~E.~Chudakov, 
Texas Symposium on Relativistic Astrophysics, (14th, Dallas, TX, Dec. 11-16, 1988) {\bf 571} 577 (1989)

\bibitem{Agl91A} 
M.~Aglietta, A.~Castellina, W.~Fulgione {\em et al.}, 
Il Nuovo Cimento C {\bf 14} 171 (1991) 

\bibitem{Agl91B} 
M.~Aglietta, A.~Castellina, W.~Fulgione {\em et al.}, 
Il Nuovo Cimento B {\bf 106} 1257 (1991 

\bibitem{Gal16} 
P.~Galeotti, G.~Pizzella, 
Eur. Phys. J. C {\bf 76} 426 (2016)

\bibitem{Agl86} 
M.~Aglietta, G.~Badino, G.~F.~Bologna {\em et al.}, 
Il Nuovo Cimento C {\bf 9} 185 (1986)

\bibitem{Ale89} 
E.~N.~Alekseev, L.~N.~Alekseeva, V.~N.~Zakidyshev, 
Pis'ma Zh. Eksp.Teor.Fiz. {\bf 49} N9 480 (1989)

\bibitem{Saa10} 
O.~Saavedra San Martin, 
Astronomy Letters Vol.{\bf 36} N7 467 (2010)

\bibitem{Bon78} 
P.~Bonifazi, V.~Ferrari, S.~Frasca {\em et al.}, 
Il Nuovo Cimento C {\bf 1} 465 (1978)

\bibitem{Ezr70} 
D.~H.~Ezrow, N.~S.~Wall, J.~Weber, and G.~B.~Yodh, 
Phys. Rev. Lett. {\bf 24} 945 (1970)

\bibitem{Mur79} 
V.~S.~Murzin {\it Introduction to Physics of Cosmic Rays}, Moscow, Atomizdat (1979) p. 304 (in russian)

\bibitem{Ama86} 
E.~Amaldi and G.~Pizzella, 
Il Nuovo Cimento C {\bf 9} 612 (1986)

\bibitem{Fal01} 
E.~D.~Falkenberg,
Apeiron {\bf 8} 32 (2001)

\bibitem{Jav10} 
D.~Javorsek~II, P.~A.~Sturrock, R.~N.~Lasenby {\em et al.}, 
Astropart. Phys. {\bf 34} 173 (2010)

\bibitem{Jen09} 
J.~H.~Jenkins, E.~Fischbach, 
Astropart. Phys. {\bf 31} Issue 6 407 (2009)

\bibitem{Jen10} 
J.~H.~Jenkins, D.~W.~Mundy, E.~Fischbach, 
Nuclear Instruments and Methods in Physics Research Section A: Accelerators, Spectrometers, Detectors and Associated Equipment {\bf 620} Issues 2-3 332 (2010)

\bibitem{Par11} 
A.~Parkhomov,
Journal of Modern Physics {\bf 2} N11 1310 (2011)

\bibitem{Moh16} 
T.~Mohsinally, S.~Fancher, M.~Czerny, E.~Fischbach, J.~T.~Gruenwald, J.~Heim, J.~H.~Jenkins, J.~Nistor, D.~O'Keefe, 
Astropart. Phys. {\bf 75} 29 (2016)

\bibitem{Alb86} 
D.~E.~Alburger, G.~Harbottle, and E.~F.~Norton, 
Earth and Planetary Science Letters {\bf 78} Issues 2-3 168 (1986)

\bibitem{7a} 
A.~Aoki, and J.~Soda, 
Phys. Rev. {\bf D93} 083503 (2016) 

\bibitem{Fis09} 
E.~Fischbach, J.~B.~Buncher, J.~T.~Gruenwald {\em et al.}, 
Space Sci Rev. {\bf 145} 285 (2009) 

\bibitem{37}
C.~Scarlett, B.~Freeman, E.~Fischbach, et al., to be published.

\bibitem{Ree74} 
M.~Rees, J.~A.~Wheeler, and R.~Ruffini, 
{\it Black Holes, Gravitational Waves and Cosmology}, 
Gordon and Breach N.Y. (1974)

\bibitem{Ims95} 
V.~S.~Imshennik, 
Space Sci. Rev. {\bf 74} 325 (1995) 

\end{thebibliography}
\end{document}